\documentclass[letterpaper]{article}
\usepackage{aaai}
\usepackage{times}
\usepackage{helvet}
\usepackage{courier}
\usepackage{graphicx}
\usepackage{amsmath}
\usepackage{amsfonts}
\usepackage{amsthm}
\usepackage{natbib}
\DeclareMathOperator*{\argmin}{arg\,min}

\title{To Recommend or Not?\\A Model-Based Comparison of Item-Matching Processes}
\author{Serina Chang\\
Stanford University\\
\texttt{serinac@cs.stanford.edu}
\And 
Johan Ugander\\
Stanford University\\
\texttt{jugander@stanford.edu}
}

\begin{document}
\nocopyright
\maketitle
\begin{abstract}
Recommender systems are central to modern online platforms, but a popular concern is that they may be pulling society in dangerous directions (e.g., towards filter bubbles). 
However, a challenge with measuring the effects of recommender systems is how to compare user outcomes under these systems to outcomes under a credible counterfactual world without such systems.
We take a model-based approach to this challenge, introducing a dichotomy of process models that we \textit{can} compare: (1) a ``recommender'' model describing a generic item-matching process under a personalized recommender system and (2) an ``organic'' model describing a baseline counterfactual where users search for items without the mediation of any system.
Our key finding is that the recommender and organic models result in dramatically different outcomes at both the individual and societal level, as supported by theorems and simulation experiments with real data. 
The two process models also induce different trade-offs during inference, where standard performance-improving techniques such as regularization/shrinkage have divergent effects. Shrinkage improves the mean squared error of matches in both settings, as expected, but at the cost of less diverse (less radical) items chosen in the recommender model but more diverse (more radical) items chosen in the organic model.
These findings provide a formal language for how recommender systems may be fundamentally altering how we search for and interact with content, in a world increasingly mediated by such systems.
\end{abstract}

\newcommand{\ju}[1]{\textcolor{blue}{#1}}
\newcommand{\jucom}[1]{\textcolor{blue}{[#1]}}
\newcommand{\syc}[1]{\textcolor{magenta}{#1}}
\newcommand\syccom[1]{\textcolor{magenta}{[SC: #1]}}

\newcommand\mus{\mu_{\text{user}}}
\newcommand\mm{\mu_{\text{item}}}
\newcommand\mn{\mu_{\text{noi}}}
\newcommand\Su{\Sigma_{\text{user}}}
\newcommand\Sm{\Sigma_{\text{item}}}
\newcommand\Si{\Sigma_i}
\newcommand\Sr{\Sigma_r}
\newcommand\su{\sigma^2_{\text{user}}}
\newcommand\sm{\sigma^2_{\text{item}}}
\newcommand\si{\sigma^2_i}
\newcommand\sr{\sigma^2_r}

\newcommand\X{\mathcal{X}}
\newcommand\Y{\mathcal{Y}}
\newcommand\zij{z_j^{(i)}}  
\newcommand\zri{z_i^{(r)}}  

\newcommand\hij{\hat{y}_j^{(i)}} 
\newcommand\hijmle{\hat{y}_j^{(i, \text{MLE})}}
\newcommand\hijmap{\hat{y}_j^{(i, \text{MAP})}}
\newcommand\hijshr{\hat{y}_j^{(i, S)}}
\newcommand\hri{\hat{x}_i^{(r)}}
\newcommand\hrimle{\hat{x}_i^{(r, \text{MLE})}}
\newcommand\hrimap{\hat{x}_i^{(r, \text{MAP})}}

\newcommand\ki{k_i}
\newcommand\kio{\ki^{\text{org}}}
\newcommand\kiomle{\ki^{(\text{org})}}
\newcommand\kiomap{\ki^{(\text{org}, \text{MAP})}}
\newcommand\kir{\ki^{\text{rec}}}
\newcommand\kirmle{\ki^{(\text{rec})}}
\newcommand\kirmap{\ki^{(\text{rec}, \text{MAP})}}

\newcommand\yki{y_{k_i}}
\newcommand\ykio{\yki^{\text{org}}}
\newcommand\ykiomle{\yki^{(\text{org})}}
\newcommand\ykiomap{\yki^{(\text{org}, \text{MAP})}}
\newcommand\ykir{\yki^{\text{rec}}}
\newcommand\ykirmle{\yki^{(\text{rec})}}
\newcommand\ykirmap{\yki^{(\text{rec}, \text{MAP})}}
\newcommand\Yk{\Y_k}
\newcommand\Yko{\Yk^{\text{org}}}
\newcommand\Ykomle{\Yk^{(\text{org})}}
\newcommand\Ykomap{\Yk^{(\text{org}, \text{MAP})}}
\newcommand\Ykr{\Yk^{\text{rec}}}
\newcommand\Ykrmle{\Yk^{(\text{rec})}}
\newcommand\Ykrmap{\Yk^{(\text{rec}, \text{MAP})}}

\newcommand\Eyki{\mathbb{E}[\yki]}
\newcommand\Eykio{\mathbb{E}[\ykio]}
\newcommand\Eykiomle{\mathbb{E}[\ykiomle]}
\newcommand\Eykiomap{\mathbb{E}[\ykiomap]}
\newcommand\Eykir{\mathbb{E}[\ykir]}
\newcommand\Eykirmle{\mathbb{E}[\ykirmle]}
\newcommand\Eykirmap{\mathbb{E}[\ykirmap]}
\newcommand\Vyki{\text{Var}[\yki]}
\newcommand\Vykio{\text{Var}[\ykio]}
\newcommand\Vykiomle{\text{Var}[\ykiomle]}
\newcommand\Vykiomap{\text{Var}[\ykiomap]}
\newcommand\Vykir{\text{Var}[\ykir]}
\newcommand\Vykirmle{\text{Var}[\ykirmle]}
\newcommand\Vykirmap{\text{Var}[\ykirmap]}

\newcommand\lossi{l_i(\X, \Yk)}
\newcommand\El{\mathbb{E}[(\yki - x_i)^2]}
\newcommand\Elo{\mathbb{E}[(\ykio - x_i)^2]}
\newcommand\Elomle{\mathbb{E}[l_i^{\textrm{(org)}}]}
\newcommand\Elomap{\mathbb{E}[l_i^{\textrm{(org, MAP)}}]}
\newcommand\Elr{\mathbb{E}[(\ykir - x_i)^2]}
\newcommand\Elrmle{\mathbb{E}[l_i^{\textrm{(rec)}}]}
\newcommand\Elrmap{\mathbb{E}[l_i^{\textrm{(rec, MAP)}}]}

\newcommand\LimEykiomle{\frac{\sm}{\sm + \si} x_i}
\newcommand\LimVykio{(\frac{1}{\sm} + \frac{1}{\si})^{-1}} 
\newcommand\BigLimVykio{\left(\frac{1}{\sm} + \frac{1}{\si}\right)^{-1}}

\newcommand\VYk{\text{Var}[\Yk]}
\newcommand\VYko{\text{Var}[\Yko]}
\newcommand\VYkomle{\text{Var}[\Ykomle]}
\newcommand\VYkomap{\text{Var}[\Ykomap]}
\newcommand\VYkr{\text{Var}[\Ykr]}
\newcommand\VYkrmle{\text{Var}[\Ykrmle]}
\newcommand\VYkrmap{\text{Var}[\Ykrmap]}

\newcommand\EVYk{\mathbb{E}[\VYk]}
\newcommand\EVYko{\mathbb{E}[\VYko]}
\newcommand\EVYkr{\mathbb{E}[\VYkr]}
\newcommand\VX{\text{Var}[\X]}

\newcommand\EVX{\mathbb{E}[V_x]}
\newcommand\mse{\mathrm{MSE}(x_i)}

\newcommand\eij{\epsilon_{ij}}
\newcommand\ei{\epsilon_{i}}
\newcommand\Fnoi[2]{F_{\text{noi}}(#1, #2)}
\newcommand\Fit[2]{F_{\text{it}}(#1, #2)}
\newcommand\closest[3]{\Pr(#1\text{ is closest}|#2, #3)}

\newcommand\rij{r_{ij}}
\newcommand\hrij{\hat{r}_{ij}}

\newcommand\Eski{\mathbb{E}[\sigma^2(y_{k_i})]}
\newcommand\Esx{\mathbb{E}[\sigma^2(x)]}
\newcommand\Esk{\mathbb{E}[\sigma^2(y_k)]}
\newcommand\Eyk{\mathbb{E}[y_k]}

\noindent
Personalized recommender systems guide the modern online experience. 
These systems recommend movies and music on content platforms \citep{nguyen2014movies,anderson2020spotify,holtz2020engagement}, suggest friendships and groups on social networking sites \citep{gupta2013wtf,kloumann2014community}, select advertisements for target audiences \citep{lambrecht2019gender}, and filter news to consumers \citep{das2007google}. 
As recommender systems increasingly shape the content we consume, we have become more critical of their potential for unintended societal consequences. For example, one concern that has received considerable attention by academics (e.g., \citep{nguyen2014movies,flaxman2016bubble,rastegarpanah2019fire}) and the public (e.g., \citep{nyt2011echo}) alike is that these systems may be pulling society towards ``filter bubble'' dynamics, where individuals become 
isolated from viewpoints besides their own \citep{pariser2011bubble}. However, a key challenge with measuring the true effects of recommender systems is that we only observe user outcomes under recommender systems, but we are missing a credible assessment of user outcomes in the absence of these systems. To understand definitively whether recommender systems have a significant impact on filter bubbles, or any other societal phenomena, we need a systematic way to compare these two scenarios and their outcomes against each other.

To address this need, we introduce a dichotomy of process models that describes a world with and a world without recommender systems. Consider the example of a user who is new to a movie distribution platform, and seeking to watch one movie for the night. Under our \textit{organic model}, the user searches for movies ``organically'' without the mediation of any recommender system. She wants to watch the movie that best fits her interests, but she does not know the contents of any of the candidate movies. During this search process she has access only to noisy signals of each movie (e.g., by watching trailers or reading reviews), and based on these noisy signals, she must estimate each movie's content. To complete the matching process, she chooses to watch the movie that she estimates best matches her own interests. 

Compare this first model against the case where the platform has a personalized recommender system. Under our \textit{recommender} model, the situation is now flipped: whereas the user knew her own interests well but previously had to estimate the contents of the movies, now the system knows the movies on the platform well, but must estimate the user's interests. In a similar manner, the system gathers a noisy sample from the user (e.g., by asking her to name one movie she recently enjoyed). Based on this noisy sample, the system then estimates her interests and recommends her the movie it believes is the best match.
In this dichotomy, the organic model functions as a baseline against which the matching behavior of a recommender system can be compared.

What connects these two settings is the act of trying to match users to items with only noisy information about one side; what differs, however, is which information is known and who is doing the matching (the user themselves vs.\ the recommender system).
We find that this difference in perspective between the two models results in significant differences in outcomes.
We characterize these differences through the lenses of both (1) individual metrics (what is the expected loss for a given user? Does this differ across users?) and (2) population metrics (what is the average user loss? Which items tend to be selected overall?). 
Furthermore, we document the notably varying effects of regularization in each of the models, where the users or system apply shrinkage to their estimates. We will see that shrinkage can reduce average user loss in both models, but it also introduces new effects that cause the models to diverge further.

Thus, even in these distilled settings, the intersection of our models, metrics, and algorithm design decisions (such as shrinkage) creates a rich environment in which we can investigate the effects of recommender systems. We summarize our contributions as:
\begin{enumerate}
    \item A framework of two contrasting models that capture organic search and recommended item-matching as symmetric, comparable processes;
    \item Theorems proving key differences between the models from both the individual and population perspectives;
    \item Simulations demonstrating that our findings translate from the theorem settings to realistic data (MovieLens).
\end{enumerate}
Collectively, our work lays out a formal model-based language for how the effect of recommender systems can be meaningfully analyzed relative a counterfactual world without such systems. 



\section{Related Work}
The widespread adoption of personalized recommender systems has led to diverse investigations of the potential societal consequences of these systems. One body of literature tackles bias in recommender systems \citep{chen2020survey,engin2013bias}, examining how they may 
systematically underrepresent minority views \citep{stoica2019hegemony}, 
serve predictions of uneven quality  across user groups \citep{yao2017parity}, or fail to recommend valuable items to certain users, e.g., showing job opportunities in STEM fields to fewer women than men \citep{lambrecht2019gender}.
Empirical observations, however, have often been mixed in nature, e.g., documenting how systems sometimes favor long-tail items \citep{fleder2009blockbuster,brynjolfsson2011goodbye} or (over-)favor popular items \citep{abdollahpouri2017popularity}. 

Another topic of societal concern is the possibility that personalized recommendations are pushing individuals into ``filter bubbles'' \citep{pariser2011bubble}. Social media users are known to selectively share content and connect to friends who agree with their existing opinions \citep{an2014selective,garimella2018bipartisan}, and there is some evidence that recommender systems exacerbate this dynamic of ideological segregation \citep{bakshy2015facebook,flaxman2016bubble}. 
Recommender systems seem to have a narrowing effect in other contexts as well: e.g., the sets of movies \citep{nguyen2014movies} and songs \citep{anderson2020spotify} recommended to users tend to be less diverse than the content that users find on their own. 
However, some studies have pointed out that the role of recommender systems is modest compared to the impact of user choice (e.g., whether to click on a recommended story) 
on narrowing consumption diversity \citep{bakshy2015facebook}. Other works have made the case that recommender systems actually \textit{increase} diversity in exposure \citep{flaxman2016bubble} and widen users' interests \citep{hosanagar2013tribes}. 


Part of the reason why it is so challenging to measure the true effects of recommender systems---and perhaps why empirical studies have not been able to reach a consensus---is that we typically only observe user outcomes under the recommender system, but we cannot assess user outcomes in the absence of these systems. Claims, e.g., that users are entering filter bubbles, must be understood relative to some counterfactual baseline. Thus, it is often helpful to design models that enable us to ``observe`` and compare these hypothetical realities. For example, \citet{dandekar2013polarization} show with a model of opinion formation that if users are biased in how they process evidence (drawing undue support for their initial opinion), recommender systems can cause a polarizing effect on users' opinions. \citet{stoica2019hegemony} show under their model that recommender systems accelerate hegemonic dynamics on social media (i.e., when a single viewpoint receives undue attention). 
\citet{perra2019opinion} model the impact of algorithmic personalization on user opinions, 
although the effect also depends on the structure of the users' social network, 
As a final example, \citet{chaney2018confounding} use models to analyze the consequences of feedback loops, when the recommender system is trained on data that was influenced by the system.

Using models to study the impact of recommender systems can be powerful when it is difficult to observe ``what-ifs'' in real life, e.g., what if we removed the feedback loop, what if recommendations were not personalized. 
In the present work, we take a model-based approach to ask arguably the most fundamental ``what-if'' question about recommender systems: what if there were no recommender system at all? 
We analyze our models using first-principles measures such as the expected item match for each user, their expected loss, and the variance over the population of matched items (Section \ref{sec:metrics}). These measures form the building blocks for many of the downstream phenomena of interest related to recommender systems, including polarization, filter bubbles, user satisfaction, user retention, and bias and fairness. Implicitly embedded in these more complex processes are the metrics we study, and thus, our work forms a foundation upon which future researchers can build. 

\section{Models and Metrics}
Our models describe two contrasting processes through which users could be matched with items of choice, e.g., movies, news articles, or consumer products. In Section \ref{sec:model-definition}, we introduce the notation and formal dynamics of our models. In Section \ref{sec:metrics}, we define the individual and population-level metrics we will use to compare the models' outcomes.

\subsection{Model definition} \label{sec:model-definition}
In both models, we will have $m$ users and $n$ items. Each user and each item has a latent position; for example, this could represent movie attributes in the movie context or ideology in the context of news articles. We denote user positions as $x_i \in \mathbb{R}^{d}$ and item positions as $y_j \in \mathbb{R}^{d}$, where $d$ represents the dimensionality of the latent space. 
Overall, the set $\X = \{x_i\}_{i=1}^{m}$ represents all user positions and the set $\Y = \{y_j\}_{j=1}^n$ represents all item positions. Our analysis will make very weak distance-based assumptions about the interest/utilities of users for items: we simply assume that user $i$'s enjoyment of item $j$ is monotonically decreasing to the distance between $x_i$ and $y_j$. Thus, under both models, the decision-making agent (either the user or the recommender system) wants to match the user to the item whose position is closest to hers.

\begin{figure*}
    \centering
    \includegraphics[width=\linewidth]{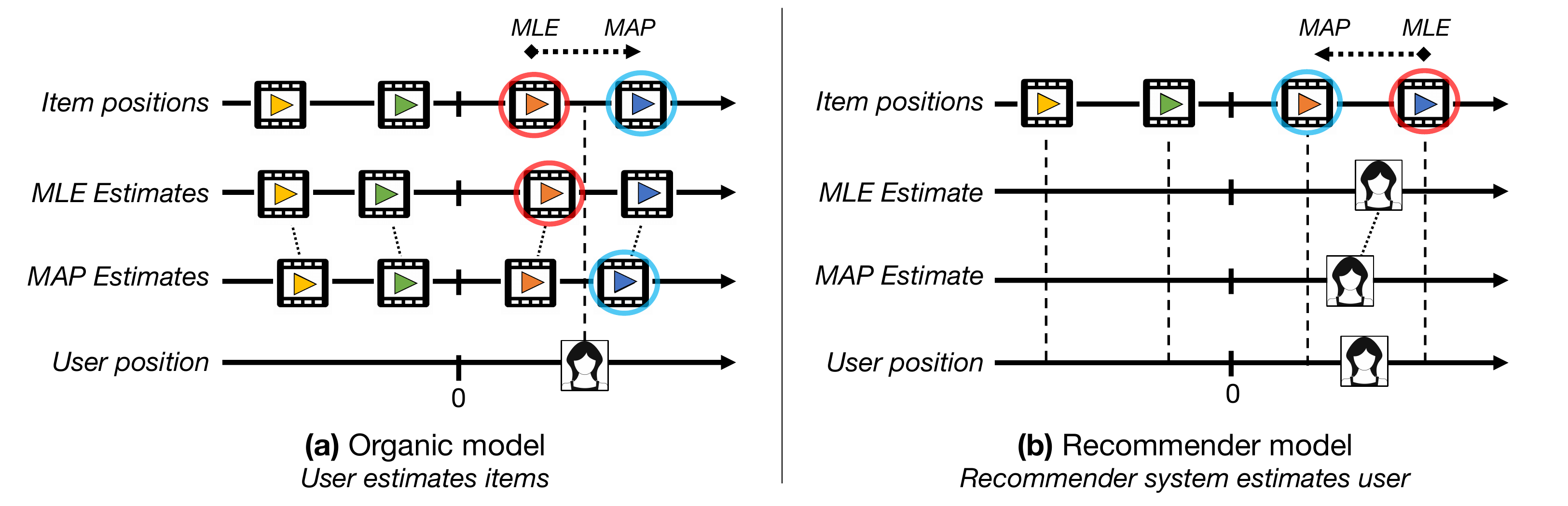}
    \caption{Model schematics. (a) In the organic model, the user samples and estimates item positions, and chooses the estimate closest to her own position. When shrinkage is applied, the user's estimates of the items shifts inward, and the user's choice switches from the inner to outer item. (b) In the recommender model, the system samples and estimates the user's position, and chooses the item closest to its estimate. When shrinkage is applied, the system's estimate of the user shifts inward, and the system's choice switches from the outer to inner item.}
    \label{fig:model}
\end{figure*}

\paragraph{Organic model.}
In this model, the user will ``organically'' search through the collection of items and choose one for herself (Figure \ref{fig:model}a). When surveying each item $j$, she only has access to a noisy sample of its true position, $y_j$. She draws her sample $\zij$ from $N(y_j, \Si)$, where $\Si$ represents the covariance in her noise.
Then, the user makes an estimate $\hij$ of item $j$'s true position.
In the simplest case, the user takes the maximum likelihood estimate (MLE) based on her single sample of $y_j$:
\begin{align}
    \hijmle = \zij.
\end{align}
We compare this MLE to an alternate case where the user applies some form of shrinkage \citep{james1961estimation} to her estimates. For example, if the user assumes a Gaussian prior on the item positions, she could take a maximum a posteriori (MAP) estimate of each, shrinking the estimate towards the item mean: 
\begin{align}
    \hijmap = (\Sm^{-1} + \Si^{-1})^{-1} (\Sm^{-1} \mm + \Si^{-1}\zij),
\end{align}
where $\mm$ and $\Sm$ are the mean and covariance of the item distribution, respectively. This style of shrinkage can also be thought of as regularization towards the item mean. In an empirical setting without a clear prior, Empirical Bayes shrinkage \citep{efron1976multivariate} would be preferred.

After surveying all $n$ items, the user will have constructed estimates for every item's position.
Since we assume utility is monotonically decreasing in distance, the user will then choose the item whose estimate is closest to her own position, $x_i$. Let $\kio$ represent the item chosen by user $i$ under the organic setting. Then,
\begin{align}
    \kio = \argmin_{j \in [n]} ||x_i - \hij||. \label{eqn:user-decision}
\end{align}
We let $\ykio$ denote the position of the chosen item $\kio$, and will later study properties of $\ykio$ as a random variable that inherits its randomness from the user's estimates, $\hij$.

\subsubsection{Recommender model.}
In this model, a recommender system takes on the burden of search instead of the user (Figure \ref{fig:model}b). When user $i$ comes onto the platform, the system gathers a noisy sample of $i$'s true position $x_i$. This sample $\zri$ is drawn from $N(x_i, \Sr)$, where $\Sr$ represents the covariance in the recommender's noise. Then, just as the user estimated item positions based on her noisy samples, the system makes an estimate $\hri$ of user $i$'s true position based on its sample of the user. Again, we first consider the case where the recommender system takes the MLE:
\begin{align}
    \hrimle = \zri.
\end{align}
We will again compare the MLE to the case where the recommender system applies shrinkage to its estimates of user positions. With a Gaussian prior, the MAP estimate of user $i$'s position is
\begin{align}
    \hrimap = (\Su^{-1} + \Sr^{-1})^{-1} (\Su^{-1} \mus + \Sr^{-1} \zri),
\end{align}
where $\mus$ and $\Su$ are the mean and covariance of the user distribution, respectively. In the absence of a prior, Empirical Bayes shrinkage would again be preferred.

As a reversal of the organic model, we assume that the recommender system has perfect knowledge of $\Y$, the set of true item positions, but can only estimate the user's true position.
Let $\kir$ represent the chosen item for user $i$ under the recommender model. Similar to the organic model, the system will choose the item whose position is closest to $\hri$:
\begin{align}
    \kir &= \argmin_{j \in [n]} ||\hri - y_j||.
\end{align}
Here we again let $\ykir$ denote the position of the matched item $\kir$, where $\ykir$ is a random variable that inherits its randomness from the system's estimates, $\hri$.

We deliberately set up the recommendation process to be as similar as possible to the organic process---the structure of the noise and logic in the choice function are identical---so that we can compare the models on the basis of who/what controls the matching and what information they have access to. We do not, for example, model the process of users reacting to recommendations. Instead, we assume that $\kir$ will be the user's match, comparing it directly to the match $\kio$ under the organic model. Furthermore, we note that both models are made up of distinct modules (sampling, estimation, choice), each of which could be swapped out for more complex processes. For example, in the organic model, one might be interested in other decision-making rules for the user, such as those that make explore/exploit trade-offs or capture risk aversion (in the case of non-uniform sample noise), or in the recommender model, the user estimation process could be derived from a specific algorithm. Thus, our models describe flexible item-matching processes that can be extended to encompass many real-world systems.

\subsection{Metrics} \label{sec:metrics}
We are interested in comparing the outcomes of these models through two lenses, at the individual and at the population levels. At the individual level, we will treat the squared distance between the user and their matched item as a user-level loss, since we model utility as monotonically decreasing in distance. Let $\Yk = \{y_{k_1}, y_{k_2}, \cdots, y_{k_m}\}$ represent the multiset of matched item positions over all users; note that $|\Yk| = |\X| = m$, and that each item position $y_j$ might appear 0 times, once, or multiple times in $\Yk$. Then, the loss for user $i$ is defined as
\begin{align}
    \lossi = ||\yki - x_i||^2.
\end{align}
As we analyze $\lossi$ in Section \ref{sec:theorems}, we will also derive $\Eyki$ and $\Vyki$ along the way; that is, the expected position and variance of the matched item for user $i$. We choose these metrics to study because they form the building blocks of many downstream phenomena of interest. For example, if the expected match for a user is not the user's own position, the matching process might eventually shift user preferences or opinions, and if different users systematically experience different losses (as we will see happens under certain settings of our models), this implies inequities in the model as it provides matches of differential quality to users on the basis of their preferences.

The population-level analysis, meanwhile, is concerned with average user loss and the overall collection of matched items. Observe that the average loss over users simply becomes the mean squared error (MSE) of the matches:
\begin{align}
    \text{MSE}(\X, \Yk) = \frac{1}{m} \sum_{i = 1}^m \lossi = \frac{1}{m} \sum_{i = 1}^m ||\yki - x_i||^2.
\end{align}
These matches are based on estimates, where shrinkage is known to reduce the MSE of an estimator. As such, there are opportunities in both models to reduce mean user loss by having the user / system apply shrinkage during estimation. However, while shrinkage might reduce MSE, that is not its only effect. In Figure \ref{fig:model}, we illustrate a basic intuition for a key result: due to the reversal in who is doing the estimating and what is being estimated, shrinkage acts as a ``diversifying'' force in the organic model, but it serves as a ``homogenizing'' force in the recommender model. We see that shrinkage in the organic model shifts the user's estimates of the items inward (i.e., towards 0), so items that are further out now have a better chance of being chosen. Meanwhile, shrinkage in the recommender model shifts the system's estimate of the {\it user} inward, so now items closer to the center are likelier to be chosen.

To quantify this intuition, we also study $\VYk$ as another population-level metric, the variance of the matched item positions $\Yk$. This variance has numerous interpretations in the real world: for example, seeing what kinds of items are being selected may guide content creators as they decide what to generate next. Furthermore, there is evidence that users adjust their preferences to better align with content that they are matched with, whether because they were persuaded by the content \citep{diehl2015persuasion}, or by the very fact that it was recommended to them \citep{summers2016ad}. Such dynamics suggest that users may become more heterogeneous if they are matched with a more diverse set of items, and more homogeneous if they are all matched to similar content. With repeated rounds of matching and opinion formation, a great level of heterogeneity in users could lead to polarization or radicalization; conversely, increasing homogeneity could result in filter bubbles and a lack of diverse perspectives for those on the platform.
\section{Model Analysis} \label{sec:theorems}
To develop a theoretical understanding of how these two models behave, we begin with simplified instances, where we assume that the user and item positions come from Gaussian distributions in a one-dimensional space. 
For now, we will assume that user and item positions are drawn independently from $N(0, \su)$ and $N(0, \sm)$, respectively (since $d=1$, we replace covariance matrices $\Sigma$ with scalar variances, $\sigma^2$). 
We also focus here on the asymptotic setting where the number of items $n$ approaches $\infty$ (in Section \ref{sec:extensions} we examine simulations with different finite values of $n$). Interestingly, we will see that even this simplified setting is sufficient to induce the phenomena we seek to understand, and that the results remain qualitatively similar when we explore non-Gaussian multidimensional user and item positions learned from data (Section \ref{sec:empirics}).

In this section, we organize our results into a series of theorems. In Theorems 1--2, we will analyze how the models differ in terms of individual metrics, by deriving 3 quantities for each model: (1) $\Eyki$, the expected item match; (2) $\Vyki$, the variance in the match; (3) $\mathbb{E}[l_i]$, the expected loss; all for any user $i$ at position $x_i$. In Theorems 3--4, we will then study how the models differ at the population level, showing that even though shrinkage can reduce MSE for both processes, a key difference is that shrinkage increases the variance of matched items under the organic model but decreases this variance under the recommender model.
\paragraph{Theorem 1.1.} In the organic MLE model as $n \rightarrow \infty$,
\begin{enumerate}
    \item $\Eykiomle \rightarrow \LimEykiomle$
    \item $\Vykiomle \rightarrow \BigLimVykio$
    \item $\Elomle \rightarrow \left(\frac{\sm}{\sm + \si}-1\right)^2 x_i^2 + \BigLimVykio$.
\end{enumerate}
\begin{proof}
In the organic MLE model, the user makes an estimate of each item position, $\hijmle = \zij$, where $\zij$ is a sample drawn from $N(y_j, \si)$. Since there are an infinite number of items and the support of the item and noise distributions is $\mathbb{R}$, then there must be at least one item estimate located at any given point in $\mathbb{R}$. Thus, when the user chooses the estimate that is closest to her own position, $x_i$, we can expect that the chosen estimate lies at $x_i$. Given the knowledge that the user's chosen item produced a sample at $x_i$, and that the chosen item's position $\ykiomle$ was drawn from $N(0, \sm)$, we can derive a closed form solution for the posterior mean and variance of $\ykiomle$ (since both distributions are Gaussian, forming a conjugate pair), forming the expressions above for $\Eykiomle$ and $\Vykiomle$. 

Finally, recall that in estimation, mean squared error can be decomposed into the sum of the squared bias and the variance of the estimator. If we view the position of the matched item, $\yki$, as an estimate for user $i$'s true position, $x_i$, then the expected squared distance between the two is simply $(\Eykiomle - x_i)^2 + \Vykiomle$, thus yielding $\Elomle$.
\end{proof}

\paragraph{Theorem 1.2.} In the organic MAP model as $n \rightarrow \infty$,
\begin{enumerate}
    \item $\Eykiomap \rightarrow x_i$
    \item $\Vykiomap \rightarrow \BigLimVykio$
    \item $\Elomap \rightarrow \BigLimVykio$.
\end{enumerate}
\begin{proof}
When the user uses MAP, her estimate of item $j$'s position becomes $\hijmap = (\sm/(\sm+\si)) \zij$. Since there are an infinite number of items, she will still choose an estimate that is equal to $x_i$; however, this means that the chosen item must have produced a sample at $((\sm + \si)/\sm)x_i$. Again, we can derive a closed form solution for the posterior distribution of the chosen item's position $\ykiomap$, knowing that it was also drawn from $N(0, \sm)$. We find that the terms cancel out, giving us $\Eykio = x_i$, and the variance does not change compared to the MLE version of the model. Since $\Eykio$ is now unbiased, the user's expected loss becomes only the variance.
\end{proof}

\paragraph{Theorem 2.1.} In the recommender MLE model as $n \rightarrow \infty$,
\begin{enumerate}
    \item $\Eykirmle \rightarrow x_i$
    \item $\Vykirmle \rightarrow \sr$
    \item $\Elrmle \rightarrow \sr$.
\end{enumerate}
\begin{proof}
In the recommender MLE model, the recommender system makes an estimate of the user's true position, $\hrimle = \zri$, where $\zri$ is a sample drawn from  $N(x_i, \sr)$. Then, the system finds the item whose position is closest to $\hrimle$. Echoing the proof in Theorem 1, if there are an infinite number of items, then there must be at least one item located at any given point in $\mathbb{R}$. Thus, the chosen item position, $\ykirmle$, must be equal to $\hrimle$, and so the distribution of $\ykirmle$ reduces to the distribution of $\hrimle$, which has an expected value $x_i$ and variance $\sr$. Since $\ykirmle$ forms an unbiased estimate for $x_i$, the expected loss for the user is again just the variance.
\end{proof}

\paragraph{Theorem 2.2.} In the recommender MAP model as $n \rightarrow \infty$,
\begin{enumerate}
    \item $\Eykirmap \rightarrow \frac{\su}{\su + \sr} x_i$
    \item $\Vykirmap \rightarrow \left(\frac{\su}{\su + \sr}\right)^2 \sr$
    \item $\Elrmap \rightarrow \left(\frac{\su}{\su + \sr} - 1\right)^2 x_i^2 + \left(\frac{\su}{\su+\sr}\right)^2 \sr$.
\end{enumerate}
\begin{proof}
In the MAP setting, the recommender model's estimate of the user's position becomes $\hrimap = (\su/(\su + \sr)) \zri$. Since there are an infinite number of items, it is still true that the chosen item's position, $\ykirmap$, will match $\hrimap$ perfectly, and so the distribution of $\ykirmap$ reduces to the distribution of $\hrimap$. Since $\hrimap$ is simply a sample drawn from $N(x_i, \sr)$ scaled by $\su/(\su + \sr)$, then $\hrimap$ must be normally distributed with mean $(\su/(\su + \sr))x_i$ and variance $(\su/(\su + \sr))^2 \sr$. This provides the expressions above for $\Eykirmap$ and $\Vykirmap$, and the expected squared loss is the sum of the squared bias, $(\Eykirmap - x_i)^2$, and the variance $\Vykirmap$.
\end{proof}

\begin{figure}[]
    \centering
    \includegraphics[width=0.8\linewidth]{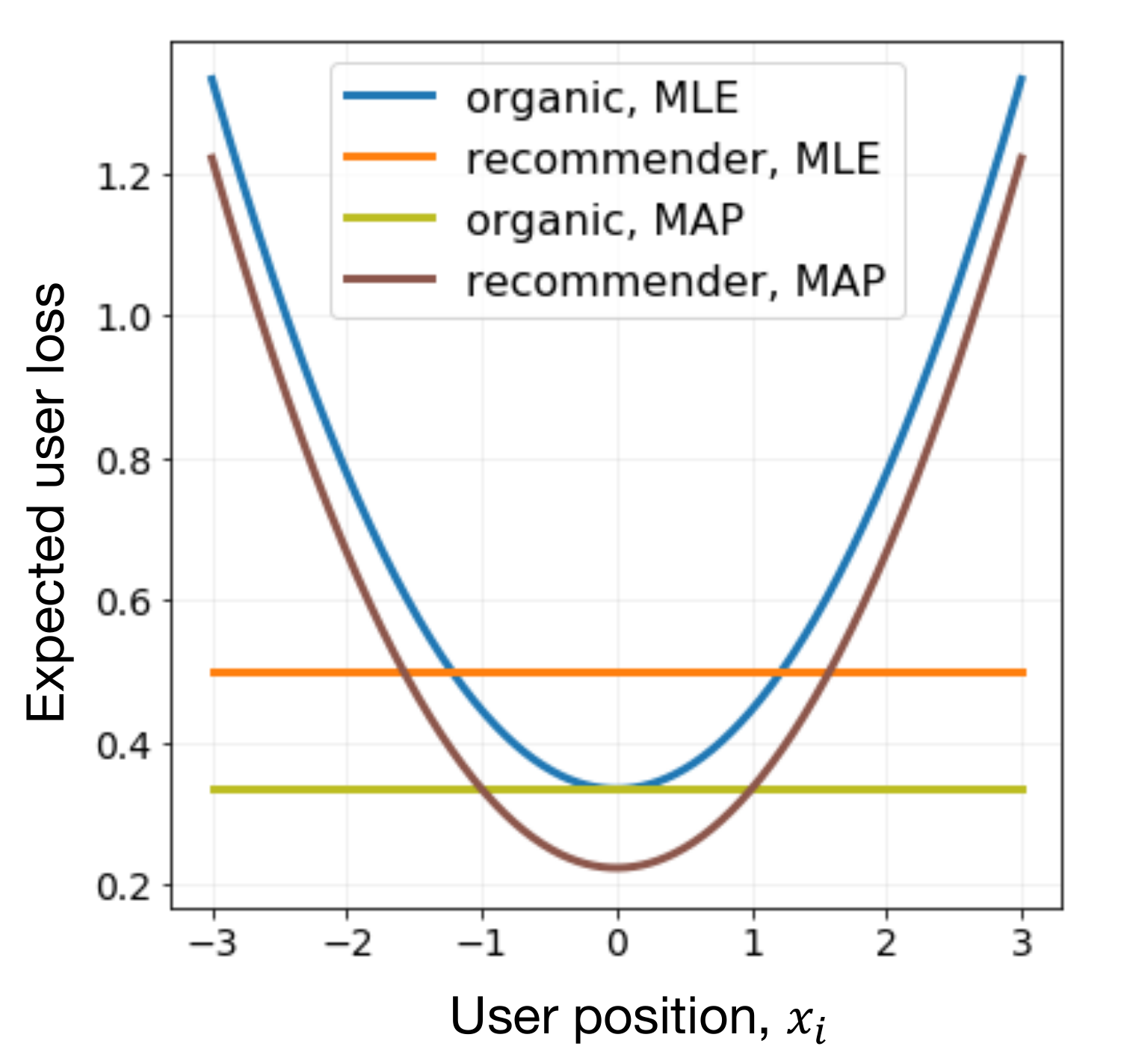}
    \caption{Expected user loss as a function of user position (Theorems 1--2); $\sm = 1$ and $\si = \sr = 0.5$.}
    \label{fig:individual_mse}
\end{figure}

Theorems 1 and 2 imply that if we treat the matched item position $\yki$ as an estimate of the user position $x_i$, then the organic MLE model forms biased matches for users, while the recommender MLE model does not. In particular, the organic MLE model is biased in that, relative to $x_i$, the expected match for user $i$ is contracted towards the center of the item distribution (Theorem 1.1). Furthermore, since the bias term grows with $x_i^2$, this means that users closer to the center will be less impacted by the bias than users who are farther out. This is a fundamental difference between the MLE versions of the processes: the organic model favors users who are closer to the center of the item distribution, while the recommender model is agnostic to user position, as we show in Figure \ref{fig:individual_mse}. 
Interestingly, the trend reverses when we switch from MLE to MAP: the organic model becomes  unbiased, and the recommender model becomes favorable to central users. Some of this effect is particular to our form of shrinkage here (namely, that the user perfectly erases the bias in the organic model by taking the MAP based on the true prior distribution of items), but in general, shrinkage will always increase bias for the recommender model, and, if the item distribution is heaviest around its mean, shrinkage will also reduce bias for the organic model.

From Theorems 1 and 2, we can also establish that switching from MLE to MAP will \textit{always} reduce mean user loss (i.e., MSE) in the organic model, and that it can sometimes reduce MSE in the recommender model. First, Theorem 1 shows that in the organic model, expected loss strictly decreases for every user (with non-zero $x_i$) from MLE to MAP; thus, the mean loss must fall as well. From Theorem 2.2, we can imagine the conditions under which MAP will achieve a lower MSE in the recommender model: if the user variance $\su$ is small, then the bias terms will be smaller because all of the $x_i^2$'s will be closer to $0$, and the relative reduction that MAP achieves on the variance, scaling by a factor of $(\frac{\su}{\su + \sr})^2$, will be larger. More generally, we can conclude that there are certainly reasons to switch from MLE to MAP for both models, which makes the consequences of the switch all the more interesting -- the individual-level consequences which we have analyzed, in terms of biased matches and favored users, and now population-level consequences, which we analyze below.

\paragraph{Theorem 3.} In the organic model as $n \rightarrow \infty$, 
\begin{enumerate}
    \item $\VYkomle \rightarrow \frac{m-1}{m}\left[(\frac{\sm}{\sm + \si})^2 \su + \LimVykio\right]$
    \item $\VYkomap \rightarrow \frac{m-1}{m}\left[\su + \LimVykio\right]$.
\end{enumerate}
\begin{proof}
In general, we can think of each chosen item position $\yki \in \Yk$ as the sum of its expected value conditioned on $x_i$, $\Eyki$, and some independent random noise $\epsilon$ with variance $\Vyki$ (which does not depend on $x_i$ in any of our models). 

Using this approach, the process of generating terms in $\Ykomle$ in the organic MLE model can be seen as first drawing a random user position $x$ from $N(0, \su)$, scaling it by $\sm/(\sm + \si)$, then adding this noise term $\epsilon$. We scale $x$ because $\Eykiomle = \sm/(\sm + \si)x_i$ (Theorem 1.1). This output becomes the sum of two random variables, where the first term is the scaled $x$ and the second is $\epsilon$. The variances of these terms are $(\sm/(\sm + \si))^2 \su$ and $\Vykio = \LimVykio$, respectively, so the expected variance of $\Ykomle$ is the sum of these expressions multiplied by $(m-1)/m$, because there are $m$ elements in $\Ykomle$.

In the organic MAP case, the process of generating terms in $\Ykomap$ can also be seen as first drawing a random user position $x$ from $N(0, \su)$, then adding a noise term $\epsilon$. We do not scale $x$ in this case, since $\Eykiomap = x_i$ (Theorem 1.2). The output is again the sum of two random variables, where the first term is $x$ and the second is $\epsilon$. Their variances are $\su$ and $\LimVykio$, respectively, so the expected variance of $\Ykomap$ is their sum scaled by $(m-1)/m$. 
\end{proof}

\paragraph{Theorem 4.} In the recommender model as $n \rightarrow \infty$, 
\begin{enumerate}
    \item $\VYkrmle \rightarrow \frac{m-1}{m}(\su + \sr)$
    \item $\VYkrmap \rightarrow \frac{m-1}{m}\left[\frac{\su}{\su+\sr} \su \right]$.
\end{enumerate}
\begin{proof}
We can use the same approach from Theorem 3 to derive $\VYk$ here. In the recommender MLE model, the process of generating terms in $\Ykrmle$ can be seen as drawing a random user position $x$ from $N(0, \su)$ then adding a noise term $\epsilon$ that has variance $\Vykirmle = \sr$ (Theorem 2.1). We do not scale $x$ here either because $\Eykirmle = x_i$. The sum of these terms' variances is $\su + \sr$, and again we scale this by $(m-1)/m$. 

In the recommender MAP model, to generate terms in $\Ykrmap$, we draw a random user position $x$ from $N(0, \su)$, scale it by $\su/(\su + \sr)$, then add a noise term $\epsilon$ (Theorem 2.2). The variance of the scaled $x$ is $(\su/(\su + \sr))^2 \su$ and the variance of $\epsilon$ is $\Vykirmap = (\su/(\su + \sr))^2 \sr$. 
Adding these terms together, simplifying, and multiplying by $(m-1)/m$ yields expression for $\VYkrmap$.
\end{proof}

\begin{figure}[]
    \centering
    \includegraphics[width=\linewidth]{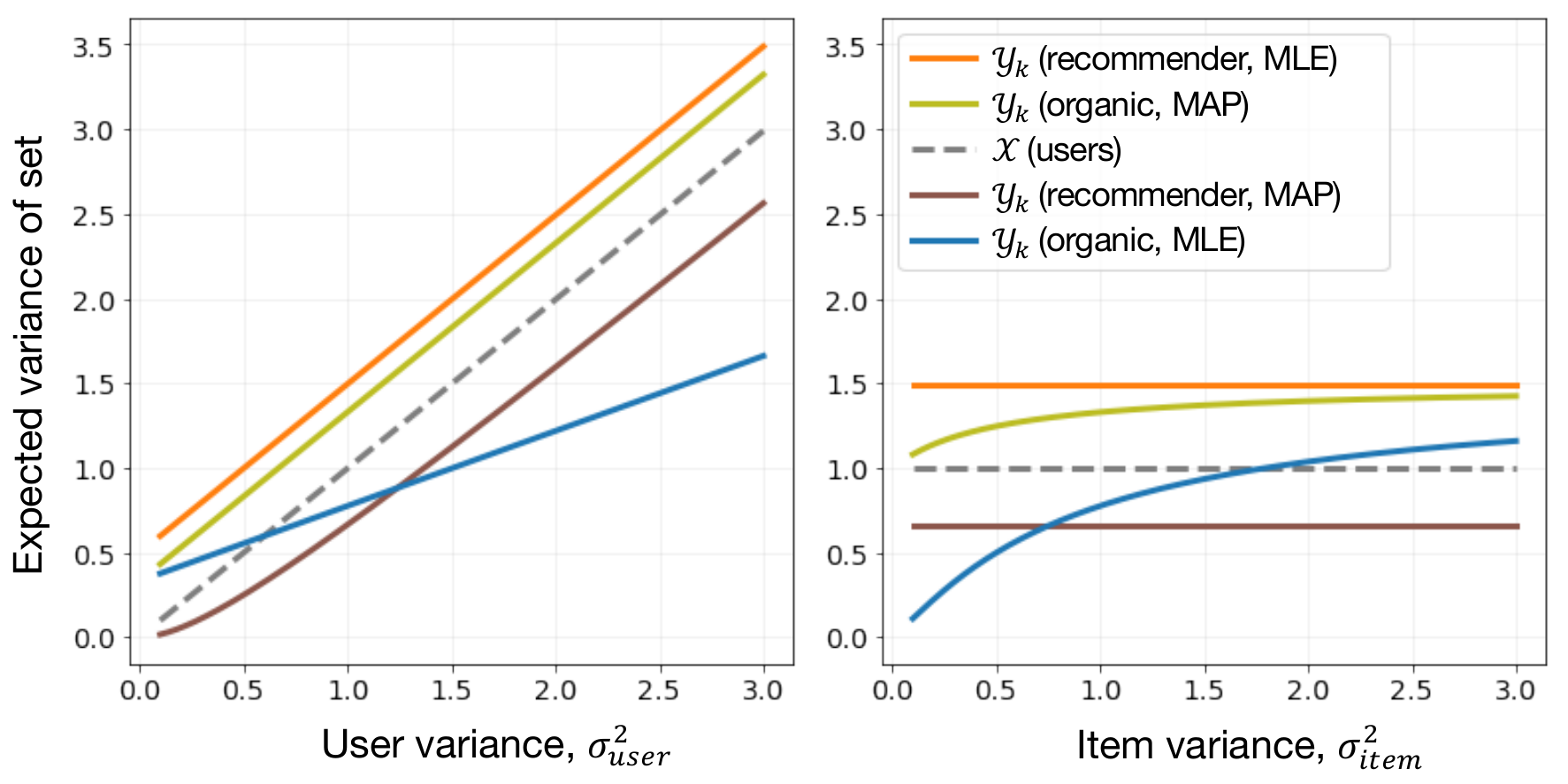}
    \caption{Variance of matched items $\VYk$ (Theorems 3--4); $\si = \sr = 0.5$ and $m=300$. On the left, we fix $\sm = 1$ and vary $\su$; on the right, we fix $\su = 1$ and vary $\sm$.}
    \label{fig:population}
\end{figure}

From Theorems 3 and 4, we can see that the variance of matched items will \textit{always} increase in the organic model when switching from MLE to MAP, but it will \textit{always} decrease in the recommender model when making the same switch. 
We demonstrate these two effects in Figure \ref{fig:population}: across different values for the user variance $\su$ and item variance $\sm$, the matched item variance for the recommender MLE model remains well above that of the recommender MAP, and the matched item variance for the organic MLE model is always below that of the organic MAP. This quantifies our earlier intuition from Figure \ref{fig:model} that shrinkage acts as a diversifying force in the organic model (increasing variance), but a homogenizing force in the recommender model (decreasing variance), and we have proven that this is always true in this setting.


\section{Simulations} \label{sec:simulations}
\begin{figure*}[tp]
    \centering
    \includegraphics[width=\linewidth]{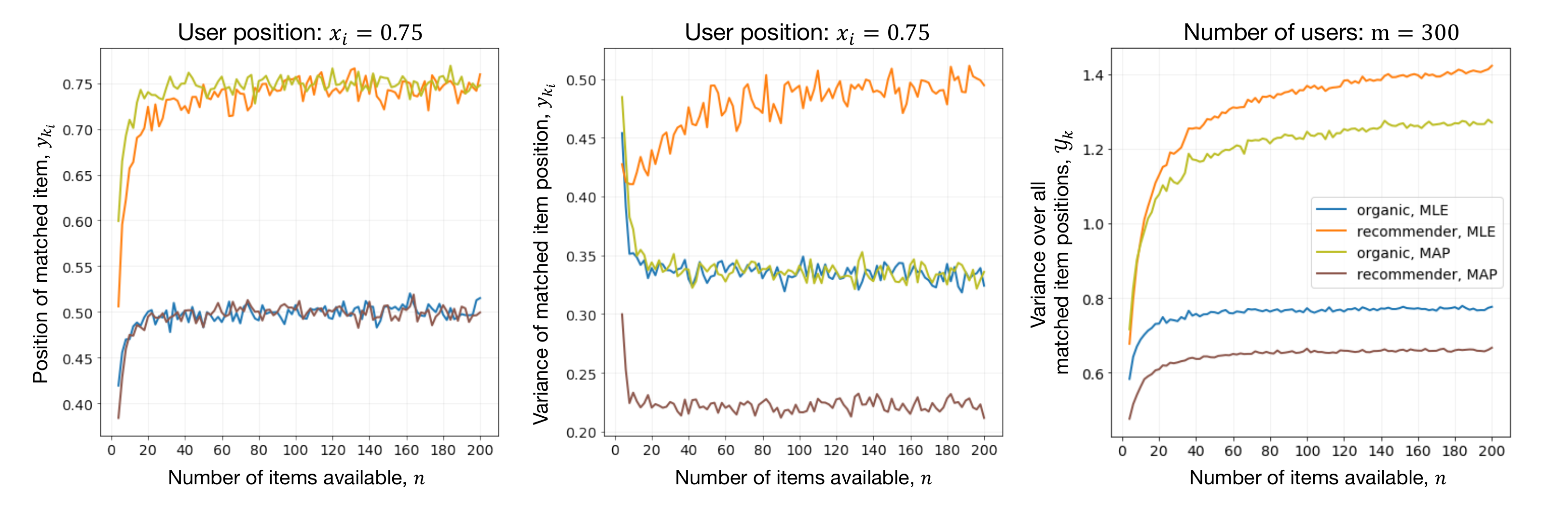}
        \vspace{-5mm}
    \caption{Simulation results over different values of $n$. As before, we set $\su = \sm = 1$, $\si = \sr = 0.5$, and $m = 300$. For the individual metrics, we run 5000 trials for each model and $n$, and evaluate the mean (left) and variance (middle) of the matched item position $\yki$ for a user positioned at $x_i = 0.75$. For the population metric, we run 500 trials for each setting and evaluate the mean variance over the matched item positions $\mathcal{Y}_k$ (right).
    }
    \label{fig:simulation_over_n}
\end{figure*}
Theorems 1--4 demonstrated that even the simplest versions of the organic and recommender processes result in vastly different matching behavior.
However, one may wonder how sensitive these results are to our assumptions (studying the asymptotic case, with single-dimensional, normally distributed user and item positions). In this section, we remove each of these assumptions, and show via simulation experiments that the key qualitative results we established in this section hold for much more realistic settings as well.

In Section \ref{sec:extensions}, we implement the single-dimensional Gaussian model described in Section \ref{sec:theorems}, and test how the individual-level and population-level metrics converge to their limits as we increase the number of available items. In Section \ref{sec:empirics}, we augment our process models to integrate multidimensional user and item embeddings learned from real ratings data.

\subsection{Exploring finite numbers of items} \label{sec:extensions}

First, in the single-dimensional Gaussian setting we analyze the behavior of our proposed metrics as the number of items $n$ grows, complementing our analytical results in Section \ref{sec:theorems}, where we focused on the asymptotic setting where $n \rightarrow \infty$. 
We consider both the MLE and MAP versions of the organic and recommender models and $n \in \{4, 6, \cdots, 200\}$. 
For the individual metrics, we simulate 5000 stochastic trials for each model and value of $n$, re-sampling per trial the positions of items and noisy samples, but fixing the location of the single hypothetical user to $x_i = 0.75$. For the population metric, we simulate 500 stochastic trials, since there is far less variance in this metric, and re-sample positions of users, items, and noisy samples per trial. 

As shown in Figure \ref{fig:simulation_over_n}, these simulations are consistent with and extend our earlier analyses. In the organic model, when the user employs the MLE, a biased match is formed and the average $\yki$ converges inward of $x_i = 0.75$; as expected, converging at $0.5$, since $\sm = 1$ and $\si$ = 0.5, and $\Eykiomle = (\sm/(\sm+\si)) x_i$ (Theorem 1). Meanwhile, the recommender MLE model forms an unbiased match, and the average $\yki$ converges to $0.75$ when the number of items $n$ reaches around 100. When MAP is used, the roles are reversed, as the organic model becomes unbiased and the recommender model becomes biased. Furthermore, every model converges to a different matched item variance: the recommender MLE model has the highest variance, followed by organic MAP, then organic MLE, then finally recommender MAP (following the order we would expect from Figure \ref{fig:population}, at $\su = \sm = 1$), and this ordering is observed when there are as few as 10 items. 

We also observe that as we increase the number of items $n$, the average $\yki$ and $\VYk$ converge to their limit from the ``inside'' (the side closer to 0). To provide some intuition for this behavior, at the core of our asymptotic analyses for $\Eyki$ is the understanding that when there are infinite items, the user will choose an estimate that lies exactly at $x_i$, regardless of the shape of the item distribution. However, with fewer (finite) items available, the more the item density influences $\Eyki$ and contracts it towards the item mean, which is 0. In the most extreme case, if there is only one item, then $\Eyki = 0$ for all users and $\VYk = 0$, because every user will be matched to the same item. Thus, as the number of items $n$ grows, $\Eyki$ and $\VYk$ converge from the side closer to 0.

\subsection{Empirical analysis with MovieLens data} \label{sec:empirics}
In this section, we further extend our simulations by incorporating multidimensional user and item positions fitted on real ratings data. We use the MovieLens 1M dataset\footnote{https://grouplens.org/datasets/movielens/1m/.}, which contains 1,000,209 ratings from approximately 6,000 MovieLens users on 3,900 movies \citep{harper2015movielens}. First, we filter the ratings matrix to only keep users that have rated at least 50 movies, then filter to keep movies with at least 50 ratings.
After filtering, we are left with a rating matrix $R \in \mathbb{R}^{m' \times n}$, where we have $m' = 4,297$ users and $n = 2,514$ movies remaining. 

\paragraph{Learning user and item distributions from data.}
We apply a collaborative filtering algorithm to $R$ in order to infer latent embeddings $\mathcal{U} = [u_i]_{i=1}^{m'}$, $u_i \in \mathbb{R}^{d}$, for each user $i$, and $\mathcal{V} = [v_j]_{j=1}^n$, $v_j \in \mathbb{R}^d$, for each movie $j$. Collaborative filtering uses the history of interactions between all users and items to infer latent representations, where users and items are encoded into low-dimensional spaces such that if a user has given an item a high rating, their representations should be ``similar'', and if a user has given an item a low rating, their representations should be further apart. Similarity can be measured in different ways (e.g., inner products, Euclidean distance); since our models take a distance-based perspective to utility, we implement a collaborative filtering procedure \citep{khoshneshin2010recsys} that embeds users and items into a unified Euclidean space where items that are closer to users are more attractive to them.

In this framework, given a user embedding $u_i$ and movie embedding $v_j$, their predicted rating $\hrij$ is
\begin{align}
    \hrij = \mu + b_i - ||u_i - v_j||^2,
\end{align}
where $\mu$ is the global average rating in $R$, $b_i$ is the user bias term capturing users that tend to rate higher or lower, and $||u_i - v_j||^2$ is the squared distance between the two embeddings. This framework naturally integrates our definition of user loss---the squared distance between a user and her matched item---since a user's predicted rating is exactly the negative squared distance, translated by $\mu + b_i$ (which remains constant per user). Furthermore, this formulation fits with the choice function in our models: just as we assume that the recommender system will choose the item that minimizes distance to its estimate of the user, that very same item here would be the one that maximizes predicted rating.  

To learn these embeddings, we define the following objective function, where we aim to minimize the regularized loss over all observed ratings in $R$:
\begin{align}
    \min_{\mathcal{U}, \mathcal{V}, \mathcal{B}} \sum_{i, j} w_{ij} [(\rij - \hrij)^2] + \lambda(||u_i - v_j||^2 + b_i^2). \label{eqn:cf_loss}
\end{align}
Here $\mathcal{B}$ represents the set of all user biases $b_i$, and $w_{ij} \in \{0, 1\}$ indicates whether the rating $\rij$ is observed in $R$. We then use gradient descent to update the latent parameters $\mathcal{U}$, $\mathcal{V}$, and $\mathcal{B}$ with respect to the regularized loss. After learning $\mathcal{U}$, i.e., the embeddings of the users in the MovieLens dataset, we sample from this empirical distribution to generate $m$ ``test'' users. This creates a new matrix of user positions, $\mathcal{X} \in \mathbb{R}^{m \times d}$, which we use in our simulations to represent unseen users. Our simulations do not rely on unseen movies, however, so we can directly set $\mathcal{Y}$, the item positions in our simulations, to $\mathcal{V}$, the movie positions directly inferred from the MovieLens dataset.

\subsubsection{Simulations with MovieLens embeddings.} 
\begin{figure}
    \centering
    \includegraphics[width=\linewidth]{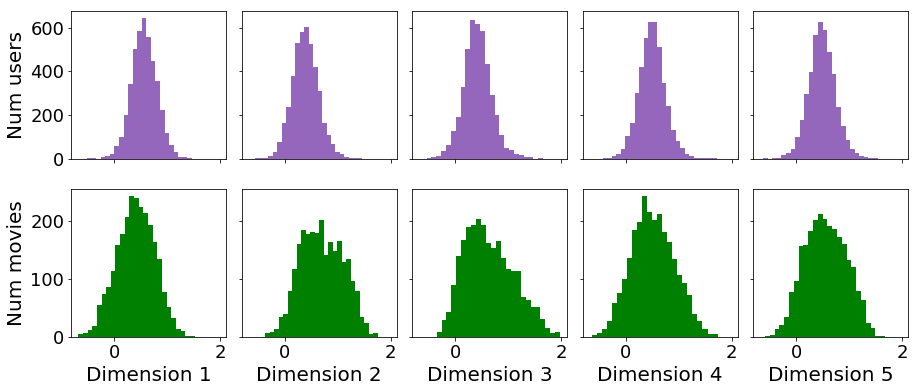}
    \caption{Distribution of learned user (top) and movie (bottom) embeddings from the MovieLens 1M dataset.}
    \label{fig:embeddings}
\end{figure}

Using this approach, we learn user and movie embeddings of dimension $d=5$ from the MovieLens dataset. In Figure \ref{fig:embeddings}, we display the learned user and movie distributions. We see that the user distribution is less dispersed than the movie distribution in each dimension, and the movie distribution is sometimes asymmetric (clearly non-Gaussian). For both embeddings, the covariance between dimensions is low.

We then simulate the organic and recommender models with $m = 300$ users and $n = 2,514$ movies (the number of movies in the filtered MovieLens dataset). For the organic model, we fix each user's noise covariance $\Si$ to $0.5 \cdot \Sm$, where $\Sm$ was the empirical covariance fitted on $\mathcal{Y}$, the movie embeddings. Similarly, for the recommender model, we fixed the system's noise covariance $\Sr$ to $0.5 \cdot \Su$, where $\Su$ was the empirical covariance fitted on $\mathcal{X}$, the user embeddings. We scaled the user / movie covariance in this way, as opposed to using spherical noise, so that the size of the noise per dimension would scale with the variance of the estimated population in that dimension, which we believed was more realistic. When shrinkage is applied within each model, we no longer have the user or system construct MAP estimates with Gaussian priors: the prior distribution is neither Gaussian nor known. Instead, we implemented shrinkage parameterized by a scalar $\alpha \in [0, 1]$, which interpolates between the MLE estimate and the mean over all MLE estimates. That is, user $i$'s shrunken estimate $\hijshr$ for movie $j$ becomes
\begin{align}
\label{eq:js}
    \hijshr = (1 - \alpha)\hijmle + \alpha (\frac{1}{n} \sum_{k=1}^n \hat{y}_{k}^{(i, \text{MLE})}),
\end{align}
and the recommender system's shrunken estimate of user $i$ is the equivalent interpolation between $\hrimle$ and the mean MLE estimate over all users. This family of estimators generalizes James--Stein shrinkage and Empirical Bayes estimators, which correspond to specific recipes for choosing $\alpha$ (or $\alpha_i$, different for each user). 

\begin{figure}
    \centering
    \includegraphics[width=\linewidth]{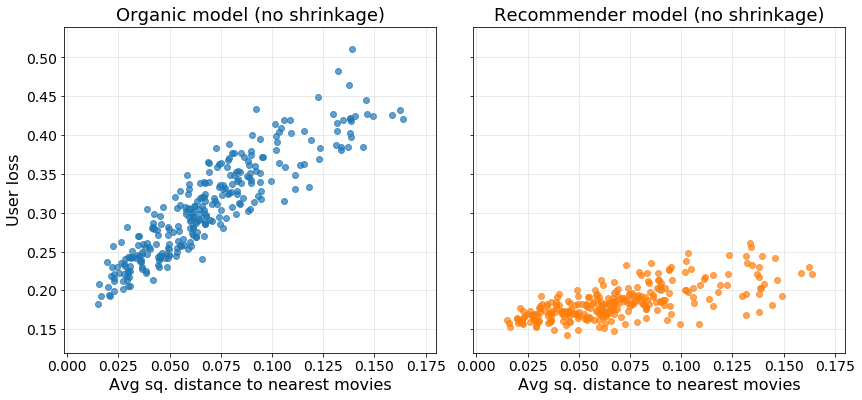}
    \caption{Individual-level results from MovieLens experiments. We plot each user's average squared distance to their 10 nearest movies (as a measure of centrality) vs. the user's loss, averaged over 500 trials. In order to see data points more clearly, we truncated outliers beyond the 95th percentile of the x-axis (showing 285 out of 300 users).}
    \label{fig:movielens-individual}
\end{figure}

\begin{figure}
    \centering
    \includegraphics[width=\linewidth]{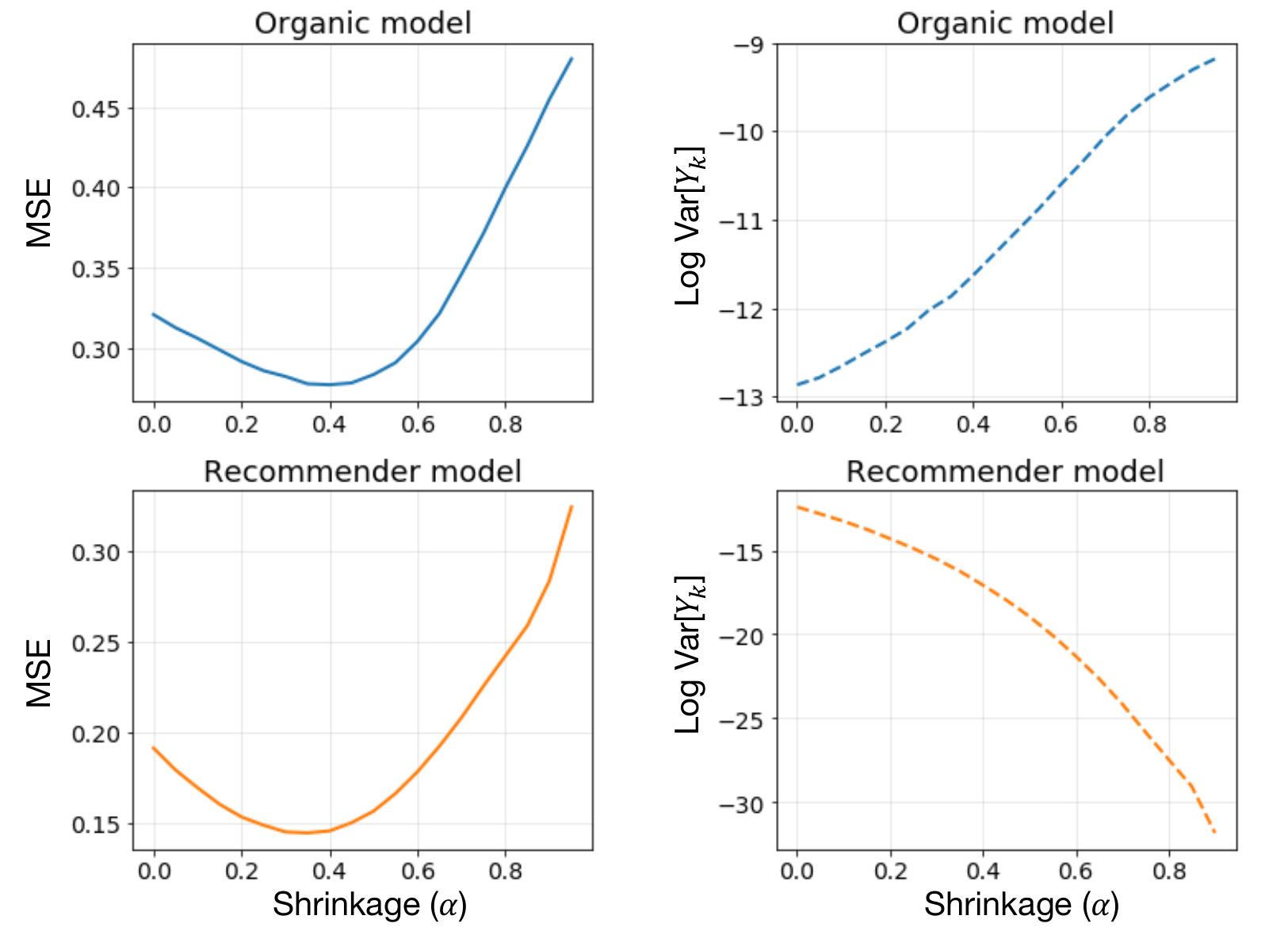}
    \caption{Population-level results from MovieLens experiments. We assess the effect of varying the shrinkage parameter $\alpha$ on the MSE and variance of matched item positions $\VYk$, running 500 trials for each model and value of $\alpha$.}
    \label{fig:movielens-population}
\end{figure}


Despite the lopsided movie distribution and expanded number of dimensions, we find that the key qualitative results from our theoretical analyses hold. 
In Figure \ref{fig:movielens-individual}, we see that the organic MLE model continues to strongly favor more central users, compared to the recommender MLE model, which is much more even-handed across user positions. 
In our earlier setting with items drawn from $N(0, \sm)$, the centrality of a user could be summarized by the absolute value of $x_i$, but here we need a more empirical measure; for example, we use the user's average squared distance to their 10 nearest movies (where higher average distance corresponds to lower centrality). 

Secondly, we simulate different amounts of shrinkage ($\alpha \in \{0.05, 0.1, \cdots, 0.95\}$) and evaluate the impacts on our population-level metrics, the mean user loss (MSE) and the variance of matched item positions $\VYk$. In multiple dimensions, we compute the generalized variance as the product of the eigenvalues of the covariance matrix of $\Yk$; we log-transform this quantity to make it more interpretable. As we increase the amount of shrinkage, at first this improves MSE for both models, but eventually MSE begins to increase; the optimal level of shrinkage seems to fall around $\alpha \approx 0.4$ for both models. However, even though the MSE curves look similar, the $\VYk$ curves completely diverge: the more we increase $\alpha$, the more $\VYk$ grows in the organic model, and the more it shrinks in the recommender model. This matches our earlier findings that shrinkage has a ``diversifying'' effect on the organic model but a ``homogenizing'' effect on the recommenders.
\section{Conclusion}
We have introduced two contrasting models of item-matching processes: one describing a generic personalized recommender system, and the other describing a setting where users search for items organically without the mediation of any system.
In both cases, we have a decision-making agent trying to match users to items with limited information, but the difference lies in who/what is doing the matching, and what information they have access to vs.\ what is being estimated. Comparing the two, we have seen that this simple switch in perspective results in dramatic differences at both the individual and population levels.
For example, in the MLE versions of the models, the recommender model forms unbiased item matches for the users, while the organic model does not, and the organic model favors central users, while the recommender model does not. 
Applying shrinkage has notably diverging effects on the two models: while it can reduce MSE in both cases, shrinkage leads the recommender model to choose increasingly similar sets of items, while leading the organic model to diversify its selections. 
These findings, which are robust to changes in user and item distribution, single and multiple dimensions, and asymptotic and finite settings, indicate pervasive differences between recommended and organic processes of item-matching.

We have worked to study one of the core counterfactual questions about recommender systems: if there were no recommender system, how would the world be different? Our analyses provide evidence that the use of recommender systems fundamentally alters how humans interact with content. Recommender systems are usually touted as reducing search frictions in markets, but through our work we find that such reductions in search friction comes with myriad subtle trade-offs.
We hope that future studies can build upon the framework established in this paper, using our models as building blocks to analyze diverse potential effects of recommender systems, such as polarization, filter bubbles, user retention rates, or fairness across users or products. 
As a specific direction, by modeling repeated choices and incorporating learning dynamics where users learn from the items they consume \citep{diehl2015persuasion}, it should be possible to analyze how the phenomena we document here, pertaining to single choices, compound over time and affect the evolution of users' opinions and behaviors. 
Finally, we also hope in future work to tie the theory of our model to real-world experiments; for example, through randomized trials where participants are assigned to unmediated organic search versus system-mediated search processes, to understand the differences most broadly.

\section*{Acknowledgements}
This work was supported in part by funding from the Stanford Program on Democracy and the Internet and ARO MURI award \#W911NF-20-1-0252. We thank Jan Overgoor, Amel Awadelkarim, Arjun Seshadri, and Kaitlyn Zhou for helpful feedback and discussions.

\bibliography{references}
\bibliographystyle{aaai}
\end{document}